\documentstyle[12pt,epsf,epsfig]{article}

 \catcode`\@=11
\def\Bbb{\ifmmode\let\next\Bbb@\else
 \def\next{\errmessage{Use \string\Bbb\space only in math mode}}\fi\next}
\def\Bbb@#1{{\Bbb@@{#1}}}
\def\Bbb@@#1{\fam\msbfam#1}

\if@twoside
\else
\fi
\ifcase \@ptsize
\or 
\or 
\fi

\def\@citex[#1]#2{%
\if@filesw \immediate \write \@auxout {\string \citation {#2}}\fi
\@tempcntb\m@ne \let\@h@ld\relax \def\@citea{}%
\@cite{%
  \@for \@citeb:=#2\do {%
    \@ifundefined {b@\@citeb}%
      {\@h@ld\@citea\@tempcntb\m@ne{\bf ?}%
      \@warning {Citation `\@citeb ' on page \thepage \space undefined}}%
      {\@tempcnta\@tempcntb \advance\@tempcnta\@ne%
      \@tempcntb\number\csname b@\@citeb \endcsname \relax%
      \ifnum\@tempcnta=\@tempcntb 
        \ifx\@h@ld\relax%
          \edef \@h@ld{\@citea\csname b@\@citeb\endcsname}%
        \else%
          \edef\@h@ld{\ifmmode{-}\else--\fi\csname b@\@citeb\endcsname}%
        \fi%
      \else
        \@h@ld\@citea\csname b@\@citeb \endcsname%
        \let\@h@ld\relax%
      \fi}%
    \def\@citea{,\penalty\@highpenalty\,}%
  }\@h@ld
}{#1}}
\@addtoreset{equation}{section}
\def\section{\@startsection {section}{1}{\z@}{-3.5ex plus -1ex minus
 -.2ex}{2.3ex plus .2ex}{\large\bf\centering}}
\def\subsection{\@startsection{subsection}{2}{\z@}{-3.25ex plus -1ex minus
 -.2ex}{1.5ex plus .2ex}{\sc}}
\gdef\@publabel{\hfil}
\gdef\@pubdate{\null}
\gdef\@pubnumber{\null}
\gdef\@author{\null}
\gdef\@title{\null}
\gdef\@abstract{\null}
\long\def\pubdate#1{\gdef\@pubdate{#1}}
\long\def\pubnumber#1{\gdef\@pubnumber{#1}}
\long\def\publabel#1{\gdef\@publabel{#1}}
\long\def\author#1{\gdef\@author{#1}}
\long\def\title#1{\gdef\@title{#1}}
\long\def\abstract#1{\gdef\@abstract{#1}}
\def\titlerelax{
\let\maketitle\relax
\let\settitleparameters\relax
\let\consolidatetitle\relax
\let\inittitlepage\relax
\let\finishtitlepage\relax
\let\titlepagecontents\relax
\let\multithanks\relax
\let\titlebaselines\relax
\let\@makepub\relax
\let\@maketitle\relax
\let\@makeauthor\relax
\let\@makeabstract\relax
\let\@maketitlenote\relax
\let\thanks\relax
\let\titlerelax\relax}
\def\titleclean
{\gdef\@titlenote{}
\gdef\@abstract{}
\gdef\@author{}
\gdef\@title{}
\gdef\@pubdate{}\gdef\@pubnumber{}\gdef\@publabel{}
\gdef\@dpublabel{}
}

\def\@makepub{\vbox to \z@{\hbox to \textwidth{\hfill
\@publabel \hfill
\llap{\parbox[t]{0.25\textwidth}{\raggedleft\@pubnumber}}}%
\vss}}
\def\@maketitle{\vskip 60pt \begin{center}
 {\LARGE \@title \par}
 \end{center}}
\def\@makeauthor{{%
\def\and{\smallskip {\normalsize \rm and\smallskip }}
\def\And{\medskip {\normalsize \rm and\\}\medskip}
\long\def\address##1{{\def\and{\\and\\}\medskip
                                {\small \it \\##1\\}
}}
{\centering
 \vskip 3em
 \large \lineskip .75em
 \@author}
 \par}}
\def\@makedate{\vskip 1.5em
 {\raggedright \small \noindent\@pubdate \par}}
\def\@makeabstract{\vskip 1.5em
{\small
\begin{center}
{\bf ABSTRACT\vspace{-.5em}\vspace{0pt}}
\end{center}
\quotation \@abstract \endquotation}}
\def\maketitle{\titlepage
\let\footnotesize\small \setcounter{page}{0}
\@makepub
\vfil
\@maketitle
\@makeauthor
\vfil
\@makeabstract
\@thanks
\vfil
\@makedate
\if@restonecol\twocolumn \else \eject \fi
\titlerelax \titleclean
\setcounter{footnote}{0}
}

\catcode`\@=12

\begin{document}
\bibliographystyle{npb}

\newcommand{\be}{\begin{equation}}
\newcommand{\ee}{\end{equation}}
\let\b=\beta
\def\blank#1{}

\def\cdd{{\cdot}}
\def\cev#1{\langle #1 \vert}
\def\cH{{\cal H}}
\def\comm#1#2{\bigl [ #1 , #2 \bigl ] }
\def\compact{ reductive}
\def\cont{\nonumber\\*&&\mbox{}}
\def\cO{{\cal O}}
\def\cul #1,#2,#3,#4,#5,#6.{\left\{ \matrix{#1&#2&#3\cr #4&#5&#6} \right\}}

\def\dz{Dz}
\def\dz{\hbox{$d\kern-1.1ex{\raise 3.5pt\hbox{$-$}}\!\!z$}}
\def\dz{ \frac{d\!z}{2\pi i}}
\def\en{\end{equation}}
\def\enn{\end{eqnarray}}
\def\eq{\begin{equation}}
\def\eqq{\begin{eqnarray}}



\def\half#1{\frac {#1}{2}}

\def\ip#1#2{\langle #1,#2\rangle}


\def\k{k}


\def\Mf#1{{M{}^{{}_{#1}}}}
\def\mno{{\textstyle {\circ\atop\circ}}}
\def\mod#1{\vert #1 \vert}

\def\Nf#1{{N{}^{{}_{#1}}}}
\def\ni{\noindent}
\def\no{{\textstyle {\times\atop\times}}}
\def\no:#1:{\mno#1\mno}
\def\nox{{\scriptstyle{\times \atop \times}}}


\let\p=\phi
\def\posdef{ positive-definite}
\def\posdefness{ positive-definiteness}
\def\Qf#1{{Q{}^{{}_{#1}}}}
\def\Qstar{\mathop{\no:QQ:}\nolimits}

\def\reductive#1#2{#1}


\def\tr{\mathop{\rm tr}\nolimits}
\def\Tr{\mathop{\rm Tr}\nolimits}







\def\vec#1{\vert #1 \rangle}
\def\vac{\vec 0}

\def\wan{$\WA_n$ }
\def\Wb{\bar W}
\def\Wf#1{{W{}^{{}_{#1}}}}
\def\wbn{$\WB_n$ }
\def\WA{\mathop{\it WA}\nolimits}
\def\WB{\mathop{\it WB}\nolimits}
\def\WBC{\mathop{\it WBC}\nolimits}
\def\WD{\mathop{\it WD}\nolimits}
\def\WG{\mathop{\it WG}\nolimits}



\def\zz#1{(z-z')^{#1}}

\openup 1\jot

\pubnumber{DTP 98-51}
\pubdate{September 1999}

\title{{Aspects of classical backgrounds and scattering for affine Toda theory on a half-line}
}

\author{
P. BOWCOCK and M.PERKINS\thanks{ Email \tt
Peter.Bowcock@durham.ac.uk M.G.Perkins@durham.ac.uk}
\address{Dept. of Mathematical Sciences,
University of Durham,  
Durham, DH1 3LE, U.K.}
}

\abstract{In this paper we study various aspects of classical solutions to 
the affine Toda equations on a half-line with integrable boundary conditions.
We begin by finding conditions that the theory has a stable vacuum by finding a Bogomolny bound on the energy, and analysing the possible singularities of the field at the boundary. Using these constraints and extensive numerical investigations we classify the vacuum configurations and reflection factors
for $A_r^{(1)}$ Toda theories up to $r=5$.}
\maketitle

\section{Introduction}

The study of $(1+1)$-dimensional integrable theories with boundaries is of interest, not only 
because many of the physical systems which they model have such 
boundaries, but also because of the more detailed algebraic structure
which they introduce \cite{CHER,SKL}. Besides the bulk $S$-matrix, a reflection
factor $K$ is needed to describe the scattering of particles off the boundary.
In a
similar way to the case of the bulk $S$-matrices \cite{BCDS}, various algebraic
constraints allow postulates to be made for the exact forms of these
reflection matrices \cite{GZ,FK1,FK2,FK,FS,RYU}. One way these conjectures can then be checked is by using perturbation theory \cite{KIM2,EC,PB}. However, before
this is possible we must first split the field into quantum fluctuations and a classical background field configuration. The background field configuration
must satisfy one of the integrable boundary conditions. Unlike in the bulk,
this can often only be achieved by a non-trivial solution which can be
surprisingly tricky to find. Indeed such a lowest-energy background solution may
not even exist since the boundary potential may destabilise a theory which is
perfectly stable in the bulk.  

In this paper we shall look at one class of such models which have been 
recently studied --- the real coupling affine Toda theories \cite{EC3}. Some 
results on classical backgrounds and semi-classical scattering solutions
have already been obtained in a number of special cases \cite{CDRS,CDR,B,D}.
Here we present a more detailed study of such solutions, and an intriguingly
complex picture emerges.

There is one 
Toda theory for each affine Kac-Moody algebra $\hat g$. The
equation of motion is
\eq
\partial_{\mu} \partial^{\mu}\phi+{{m^2}\over \beta}\sum_{i=0}^r n_i \alpha^i
e^{\beta \alpha^i \cdot \phi}=0.
\label{eqm}
\en
Here we have used the notation that $\alpha^i$ for $i=1$ to $r$ are the 
simple roots of $g$, the finite Lie algebra associated to $\hat g$, and 
$\alpha^0=-\psi$ where $\psi$ is the highest root of $g$ 
(for untwisted algebras). The $n_{i}$ are the marks; integers such that 
\eq
\alpha_{0} + \sum_{i=1}^{r} n_{i} \alpha^{i} = 0.
\label{sum.alphas}
\en
Also we have
defined $r$ to be the rank of $g$ and introduced $m$ the mass parameter, and 
$\beta$ the coupling constant of the theory. From now on we shall arrange for 
$m=\beta =1$ for convenience. For the theory defined on the whole
line the bulk energy 
\eq
E=\int_{-\infty}^{\infty} dx \left ({1\over 2}(\partial_x \phi)^2+
{1\over 2}(\partial_t \phi)^2+\sum_{i=0}^r n_i (e^{\alpha^i \cdot \phi} -1)
\right )
\label{eq.bulk.energy}
\en
is manifestly positive for real coupling constant. However we must modify this
formula if we wish to consider the theory on the half-line, by restricting the
above integral to the physical region $x\leq 0$ and  adding a boundary potential
term. Thus the energy for the boundary theory becomes
\eq
E=\int_{-\infty}^{0} dx \left ({1\over 2}(\partial_x \phi)^2+
{1\over 2}(\partial_t \phi)^2+\sum_{i=0}^r n_i (e^{\alpha^i \cdot \phi} -1)
\right )+V_b(\phi)|_{x=0}.
\label{eq.half-line.energy}
\en
By varying this action, it can easily be shown that this corresponds to 
a theory with boundary condition
\eq
\partial_x \phi= {{-\partial V_b}\over{\partial \phi}}.
\en
\label{eq.non.Neumann}
It has been shown that to preserve the integrability of the theory
we must take the potential to have the form \cite{CDRS,BCDR,EC2,EC1}
\eq
V_b(\phi)=-2\sum_{i=0}^r A_i  \sqrt{\frac{2 n_i}{\alpha_i^2}} e^{\alpha^i \cdot \phi/2}
\label{eq.boundary.potential}
\en
 where the constants $A_i$ must either all vanish (Neumann boundary conditions)
or obey a number of restrictions which are given in \cite{BCDR}. For simply
laced theories we have that $A_i=\pm 1$.

An important point is that $A_i$ can be positive as well as negative,
and so whilst the theory in the bulk has positive energy, 
it is not clear whether the boundary potential term can destabilise the theory.
To get a large negative boundary contribution to the 
energy we need that $\alpha^i \cdot \phi$ becomes large, and this in turn produces
a positive contribution to the gradient terms in the 
energy. Which of these two competing factors wins is determined by the value of $A_i$. In this paper we determine that one needs
\eq
A_i\leq 1
\label{eq.result}
\en
for the positive bulk energy to win out and for the theory to have energy bounded below 
Our approach to finding this result will be based on generalising the
stability argument for the Sinh-Gordon model given in \cite{CDR} which was based on a Bogomolny-like bound. We review this argument and explicitly extend it to $A_2^{(1)}$ in section two of the paper. Then in section three, we find an 
implicit Bogomolny form of the energy for all the other Toda theories, and 
use this together with an analysis of the possible singularity structure of the 
solution at the boundary to prove the result (\ref{eq.result}).

In the next section of the paper we embark on a detailed study of classical
vacuum solutions to the $A_r^{(1)}$ Toda theories. This continues and 
refines the analysis in terms of tau functions given in \cite{B}. A general
analysis of the equations, incorporating the singularity analysis of section
three, is considered and a complete classification of vacuum
solutions for different boundary conditions is given for $A_r^{(1)}$
up to $r=5$. Some of the solutions in \cite{B} are excluded because of 
singularities in the physical region $x<0$, whilst a number of new `exceptional'
solutions are found.

In section five we consider the scattering solutions around the new vacuum 
solutions we have found. Again a complete classification of $K$ matrices for 
$A_r^{(1)}$ up to $r=5$ is given. The paper concludes with some conclusions.

\section{The energy on the half-line for the Sinh-Gordon theory and $A_2^{(1)}$
Toda theory}   

For the Sinh-Gordon model the parameters $A_i$ appearing in the boundary 
potential (\ref{eq.boundary.potential}) are known to be unconstrained by
integrability. However, if we would like the Hamiltonian describing the 
theory on the half-line
(\ref{eq.half-line.energy}) to have energies bounded from below, we must 
have that $A_i\leq 1$ as shown in \cite{CDR}. This result relied on 
being able to write the energy for the Sinh-Gordon theory in a 
Bogomolny form. Explicitly we have that the energy
\eq
E=\int_{-\infty}^{0} dx \left ({1\over 2}(\partial_x \phi)^2+
{1\over 2}(\partial_t \phi)^2+(e^{\sqrt{2} \phi} +e^{-\sqrt{2} \phi}-2)
\right )-2A_1e^{\phi/\sqrt{2} } -2A_0e^{-\phi/\sqrt{2} }
\label{eq.boundary.energy}
\en
can be rewritten as
\eqq
E&=&\int_{-\infty}^{0} dx \left ({1\over 2}(\partial_t\phi)^2+
{1\over 2}(\partial_x \phi- \sqrt{2}e^{\phi/\sqrt{2}}
+\sqrt{2}e^{-\sqrt{2} \phi})^2+2\partial_x(e^{\phi/\sqrt{2} }
+e^{-\phi/\sqrt{2}})
\right )\cont
-(2A_1e^{\phi/\sqrt{2} } -2A_0e^{-\phi/\sqrt{2} })_{x=0}\\
&=&\int_{-\infty}^{0} dx \left ({1\over 2}(\partial_t\phi)^2+
{1\over 2}(\partial_x \phi- \sqrt{2}e^{\phi/\sqrt{2} }+\sqrt{2}
e^{-\phi/\sqrt{2}})^2\right )\cont
-(2(A_1-1)e^{\phi/\sqrt{2} } -2(A_0-1)
e^{-\phi/\sqrt{2} })_{x=0}.
\label{eq.sg.bogomolny.energy} 
\enn

The integral is non-negative, as is the boundary term if 
$A_i\leq 1$. Furthermore we can show that this condition is {\em necessary} 
for stability by taking the field to be of the form 
\eq
e^{\phi/\sqrt{2}}={{1+De^{2x}}\over{1-De^{2x}}}
\en
where $D$ is a constant which must be taken to be less than one for
the solution to be non-singular in the region $x<0$.
This is the analytic continuation of the Sine-Gordon kink to the real
coupling theory, and it satisfies the Bogomolny equation
\eq
\partial_x \phi- \sqrt{2}e^{\phi/\sqrt{2} }+\sqrt{2}e^{-\phi/\sqrt{2}}=0
\en
so that the only contributions to the energy come from the boundary terms at
$x=0$ in (\ref{eq.sg.bogomolny.energy}). By choosing the constant $D$ to be 
close
to one, we can take the field
$\phi(0,t)$ to be as large as we like at the boundary, and hence take the
energy to be
as negative as we like if $(A_1-1)>0$. Similarly we can show that that
$A_0\leq 1$ 
is necessary for stability by considering the solution obtained by taking 
$\phi\to -\phi$ 
in the above.

We would like to generalise this result to Toda theories based on other algebras by finding a Bogomolny-like form 
\eq
E=\int_{-\infty}^{0} dx \left ({1\over 2}(\partial_x\phi_i-W_i)^2+
{1\over 2}(\partial_t \phi_i)^2+\partial_x W(\phi_i)\right ) +V_b(\phi_i)
\label{eq.bog.energy}
\en
for the energy in these cases too. Comparing this to (\ref{eq.boundary.energy}), we see that we can write the energy in this form provided that 
\eqq
\sum_{i=1}^r \frac{1}{2} W_i^2 &=& V_{bulk}=\sum_{i=0}^r n_i (e^{\alpha^i \cdot \phi} -1)\\
\partial_x W &=& \sum_{i=1}^r W_i \partial_x \phi_i
\label{eq.bog.conditions}
\enn
and from the second equation we deduce that 
\eq
W_i=\frac{\partial W}{\partial \phi_i}
\en
so that we must have the integrability condition
\eq
\frac{\partial W_i}{\partial \phi_j}=\frac{\partial W_j}{\partial \phi_i}.
\label{eq.integrable.condition}
\en

As usual, a solution of the Bogomolny equations 
$\partial_x \phi_i=W_i$ will automatically be a (static) solution to the 
Toda field equations which are the equations for the $(1+0)$-dimensional Toda
molecule:
\eq
-\phi_{xx}+{m\over \beta}\sum_{i=0}^r n_i \alpha^i
e^{\beta \alpha^i \cdot \phi}=0.
\label{eqmtm}
\en
The integrability of the Toda molecule can be used to show that any
solution to these $r$ second order equations must also satisfy $r$ first
order equations. The Toda molecule describe a finite (in fact $r$)
degrees of freedom, and integrability can be taken in the strict Liouville
sense; that is there exist $r$ conserved quantities which Poisson-commute with each other. Note that now $x$ plays
the role of time in this system. By using the equation of motion (\ref{eqmtm})
to eliminate higher derivatives, the conserved quantities which we label
$H_i$ can be written as functions of $\phi_i$ and the momenta $p_i=\partial_x
\phi_i$. Thus we have 
\eq H_i(\phi_i,p_i)=\gamma_i.
\en 
We shall be interested in solutions for which $\phi_i,p_i\to 0$ as the `time'
$x\to -\infty$, since these are the solutions which will have finite energy
in the $(1+1)$ Toda system. For such solutions $\gamma_i=H_i(0,0)$.
For instance one of the conserved charges is the energy for the $(1+0)$ system
\eq
H_1=\sum_{i=1}^r \frac{1}{2} p_i^2-V_{bulk}=\sum_{i=1}^r \frac{1}{2} p_i^2-\sum_{i=0}^r n_i (e^{\alpha^i \cdot \phi} -1)=0.
\label{eq.mol.energy}
\en
Together with the other $r-1$ equations from conserved charges, we have $r$
equations which can be used to solve for $p_i$ in terms of the $\phi_i$. We
suggestively write these solutions
as 
\eq
p_i=\partial_x \phi_i =W_i(\phi).
\label{eq.invertforp}
\en
For these equations to satisfy the criteria to be Bogomolny equations
for the Toda system we need to show that the conditions (\ref{eq.bog.conditions}) and (\ref{eq.integrable.condition}) hold.
The first condition follows immediately from (\ref{eq.invertforp}) and
(\ref{eq.mol.energy}). The second condition follows from a result in 
classical mechanics that any equations derived from Poisson commuting
quantities must also be Poisson commuting (see for instance \cite{WHI}).
Thus we must have that 
\eq
\left \{p_i-W_i,p_j-W_j\right \}=\frac{\partial W_j}{\partial \phi_i}
-\frac{\partial W_i}{\partial \phi_j}=0.
\en

This procedure can be explicitly performed in the case of the affine Toda 
theory based on the algebra $A_2^{(1)}$. For this case it is convenient
to use variables 
\eqq
u_i &=& e^{\lambda_i \cdot \phi}\\
\pi_i &=& \lambda_i \cdot \partial_x \phi
\enn
where as usual $\lambda_i$ are the highest weights of the two fundamental
representations of $A_2$ (i.e. the $3$ and the $\bar{3}$). In terms of these
variables the two conserved quantities can be written
\eqq
H_1 &=& \pi_1^2+\pi_2^2-\pi_1\pi_2-\frac{u_1^2}{u_2}-\frac{u_2^2}{u_1}-\frac{1}{u_1 u_2}+3=0\\
H_2 &=& \pi_1 \pi_2^2-\pi_1^2 \pi_2-\frac{u_2^2}{u_1}\pi_1+\frac{u_1^2}{u_2}\pi_2-
\frac{\pi_2}{u_1 u_2}+\frac{\pi_1}{u_1 u_2}=0.
\enn
These equations can be easily solved for $\pi_1$ and $\pi_2$ as follows
\eqq
\pi_1 &=& (u_1-1)\sqrt{\frac{1+u_1+u_2}{u_1 u_2}}\\
\pi_2 &=& (u_2-1)\sqrt{\frac{1+u_1+u_2}{u_1 u_2}}.
\enn
Other solutions exist but these do not correspond to real $\phi$. From this 
we see that 
\eq
W=2\sqrt{\frac{(1+u_1+u_2)^3}{u_1 u_2}}.
\en
From the Bogomolny argument it now follows that the energy of any field
configuration on the half-line is bounded by
\eqq
E &\geq& [W]^0_{-\infty}+V_b|_{x=0}\\
&=& -6\sqrt{3}+2\sqrt{\frac{(1+u_1+u_2)^3}{u_1 u_2}}
-2A_1\frac{u_1}{\sqrt{u_2}}-2A_2\frac{u_2}{\sqrt{u_1}}-2A_0\frac{1}{\sqrt{u_1 u_2}} \\
&=&  -6\sqrt{3}+2\frac{\sqrt{(1+u_1+u_2)^3}-\sqrt{u_1^3}-\sqrt{u_2^3}-1}{\sqrt{u_1 u_2}}\cont
-2(A_1-1)\frac{u_1}{\sqrt{u_2}}-2(A_2-1)\frac{u_2}{\sqrt{u_1}}-2(A_0-1)\frac{1}{\sqrt{u_1 u_2}} .
\enn
For real $\phi$, $u_i\geq 0$ and it is easy to show that the second term is bounded
below in this region. On the other hand, the last terms are also non-negative 
for 
\eq
A_i\leq 1.
\en
These conditions ensure classical stability, and indeed it can be shown that 
these conditions are also necessary.

\section{Bogomolny equations and stability for other affine Toda theories
on the half-line}

Let us now proceed to the cases of other algebras. In principle we could 
follow the same steps as for $A_2^{(1)}$ and $A_1^{(1)}$ but in practice
we are unable to invert the conserved quantities to find explicit relations
between for the momenta in terms of the fields $\phi$. However, we can use our
knowledge of (analytically continued) static solutions to the Toda equations 
to circumvent this difficulty. Any static solution must necessarily obey the conservation laws
for the Toda molecule, and so in turn must satisfy the Bogomolny equations
(\ref{eq.invertforp}), and saturate the Bogomolny bound. It follows that 
the energy density for such a solution can be written is a total derivative
of some function $W$. But it has been known for some time in the context of 
imaginary coupling Toda theories that this is indeed the case, and the 
explicit formula for $W$ is given in terms of tau functions as
\eq
W=-2  \sum_{i=0}^r \frac{2}{\alpha_i^2} \frac{\tau_i'}{\tau_i} 
\label{eq.superpotential}
\en
where
\eq
\phi=-\sum_{i=0}^r \frac{2 \alpha_i}{\alpha_i^2} \ln(\tau_i).
\label{eq.def.tau}
\en

Some clarifying remarks about the above formulae are in order. First, note 
that the equation (\ref{eq.def.tau}) cannot completely specify the $r+1$ tau
functions, since there are only $r$ components of $\phi$; indeed we can
send 
\eq
\tau_i\to \tau_i (f(x,t))^{m_i}
\label{eq.ambiguity}
\en
where 
\eq
\sum_{i=0}^r m_i \frac{2 \alpha_i}{\alpha_i^2}=0,
\en 
without affecting (\ref{eq.def.tau}). This ambiguity is partially  
removed by fixing a particular form of the resulting equations of motion
for the tau functions;
\eq 
\ddot{\tau_{i}} \tau_{i} - \dot{\tau_{i}}^{2} - \tau_{i}'' \tau_{i} + \tau_{i}'^{2} = \left(\prod_{j=0}^{r} \tau_{j}^{2\delta_{ij}-K_{ij}} - \tau_{i}^{2}\right) \frac{n_{i}\alpha_i^2}{2}
\label{eq.tau.eqm}
\en
where $K_{ij}=2\alpha_i \cdot \alpha_j/\alpha_j^2$ is the extended Cartan matrix. We are still free to do a transformation of the form (\ref{eq.ambiguity}) for any $f$ that satisfies $\partial_\mu\partial^\mu\ln(f)=0$. For time independent solutions this implies that 
\eq
f=e^{ax+b}
\en
which only adds an unimportant constant to $W$. We complete the `gauge fixing' by demanding that $\tau\to 1$ as $x\to -\infty$.
With this choice of tau functions all the static
solutions to the real Toda equations can be described in terms of $r$ parameters,
one real parameter for each node on the (non-extended) Dynkin diagram
corresponding to a
real fundamental representation, and one complex parameter for each pair of 
complex conjugate nodes. One can now consider the map between the parameters
in the tau functions and the value of the corresponding static soliton $r$-component field $\phi$ on the 
boundary which  is obtained by putting $x=0$ in (\ref{eq.def.tau}). We shall
simply assume that imposing the condition that the field is free of singularities in the physical region and and tends to zero as $x\to -\infty$
renders this map invertible so that we can write the parameters in terms of the
values of the fields of the corresponding solitons at $x=0$. In this way we 
can view $W$ as a function of $\phi$ as we did for $A_2^{(1)}$ and $A_1^{(1)}$. With this in mind, we can write the energy on the half-line of {\it any} field configuration in the form (\ref{eq.bog.energy}) so that the energy is bounded
below by
\eq
E\geq E_{bound}=W(\phi)|_{x=0}-W(\phi)|_{x=-\infty}+V_b(\phi)|_{x=0}
\label{eq.en.bound}
\en
i.e. the energy of the static soliton configuration with a value of $\phi$ 
that coincides with that of our arbitrary solution at $x=0,-\infty$ at some
point in time. 

To prove stability of Toda theory on the half-line, we must show that the 
right hand side of (\ref{eq.en.bound}) is bounded from below. This can be 
written
\eq
E_{bound}=-2 \sum_{i=0}^r\left ( \frac{2}{\alpha_i^2}\frac{\tau_i'}{\tau_i}+A_i
\sqrt{\frac{2n_i}{\alpha_i^2}}\prod_{j=0}^r |\sqrt{\tau_j^{-K_{ij}}}| \right )
\label{eq,explicit.energy}
\en
where $K_{ij}=2\alpha_i\cdot\alpha_j/\alpha_j^2$ is the extended Cartan matrix,
and we have used that $W(\phi)|_{x=-\infty}$ vanishes since $\tau_i'(x)\to 0$
as $x\to -\infty$. If the energy is not bounded below, we can tune the parameters defining the tau functions to make $E_{bound}$ arbitrarily negative.
Na\"{\i}vely this can occur two ways: either one of the tau functions become very
large, or else becomes zero. In fact the first possibility never arises, since
if one tunes the parameters to make the tau function large, the quotients
appearing in $E_{bound}$ tend to a finite limit. So if the energy is unbounded
we must have a soliton solution with one or more tau function vanishing at 
the boundary and for which $E_{bound}=-\infty$.

Our approach therefore is to consider the function $E_{bound}$ for such soliton
solutions with singularities at the boundary, and determine whether the 
residue of the pole in $x$ is positive (corresponding to infinite positive
energy or negative). If the residue is always positive (or zero) then $E_{bound}$ must be bounded below. To analyse the behaviour near the origin take
\eq
\tau_{i} = a_{i} x^{y_{i}}+O(x^{y_{i}+1})
\en
and we know that $y_i\geq 0$ and $a_i\neq 0$ for all $i$. 
Let $S\subset{0,1,...r}$ be the set of $i$ for which $y_i>0$, i.e. for 
which the corresponding tau function $\tau_i$ vanishes at $x=0$. For 
$i\in S$ the
equation of motion (\ref{eq.tau.eqm}) contains a
term of the form $x^{2y_i-2}$. Comparing coefficients we find that for 
\eqq
y_i \prod_{j=0}^r a_j^{K_{ij}}&=& \frac{n_i \alpha_i^2}{2} \label{eq.sin.conditions1} \\
\sum_{j=0}^r K_{ij}y_j &=& 2 .
\label{eq.sin.conditions} 
\enn
Note that the second equation implies that not all the tau functions can
vanish simultaneously, or in other words $S\neq \{0,1,...r\}$ since then we 
could write 
\eq
0= \sum_{i,j}n_i \frac{2\alpha_i\cdot \alpha_j}{\alpha_j^2}y_j= \sum_{i,j} n_i K_{ij} y_j
=2\sum_i n_i=2 h
\en
which is clearly not true. Also from both equations we see that for $i\in 
S$, then for $x$ close to $0$,
\eq
e^{\alpha_i\cdot \phi}=\prod_j \tau_j^{-K_{ij}}
\sim  \prod_{j=0}^r (a_jx^{y_j})^{-K_{ij}}=\frac{2 y_i}{n_i \alpha_i^2 x^2}.
\en
Thus the poles in $E_{bound}$ at $x=0$ have the form
\eq
E_{bound}\sim -\frac{2}{x}\sum_{i\in S}\left ( \frac{2y_i}{\alpha_i^2}-A_i\frac{2\sqrt{y_i}}{\alpha_i^2} \right )+O(1)
\en
where we have used that $e^{\alpha_i\cdot \phi}>0$ and $x<0$ to identify
the correct sign for the last term. Thus we see that $E_{bound}$ will be 
bounded below if and only if 
\eq A_i\leq 1,
\en
generalising the result we found explicitly for $A_2^{(1)}$ and $A_1^{(1)}$ to all other affine algebras, since
(\ref{eq.sin.conditions}) implies 
we can always arrange $y_i=1$ by arranging that only $\tau_i$ vanishes.
Note that this result implies that all simply-laced Toda theories with
integrable boundary conditions are stable, although those with $A_i=1$ are
only marginally so.
  
\section{Vacuum soliton solutions to the Toda field equations}

In this section we elaborate on the results in \cite{CDR,B} looking for 
static solutions to the $A_r^{(1)}$ Toda equations which satisfy integrable boundary equations. A large class of solutions was found in \cite{B} and
subsequently discussed in \cite{D}. These solutions typically involved
pairs of analytically continued solitons \cite{TH}. However, the analysis
could not determine whether solutions (beyond those containing a single pair
of solitons) had any singularities in the physical region $x<0$, which would render them unacceptable
since such solutions have infinite energy. Nor was it clear whether there were
other solutions to the equations, potentially with lower energy and therefore
the `true' vacuum solutions. Here we carry the analysis a little
further, utilising the singularity analysis of the previous section. Extensive
numerical analysis indicates that generally the vacuum solutions of \cite{B} 
with more than one pair of constituent solitons do develop singularities in 
the physical region and so are unacceptable. Moreover, we shall see that even for critical values of the soliton parameters where it is possible to place such singularities on the boundary, we cannot obtain acceptable vacuum solutions. However a number of exceptional
solutions are found which provide acceptable vacuum configurations for $A_r^{(1)}$ 
up to $r=5$.

Let us recall some of the notation of \cite{B}. In the language of the
tau-functions introduced in the last section, the boundary conditions can
be conveniently written (specialising to the $A_r^{(1)}$ series of affine Toda theories where $n_i=1$ and $\alpha_i^2=2$)
\be
\frac{\tau_{i}'}{\tau_{i}} +  A_{i} e^{\alpha^{i} \cdot \phi /2} = C
\label{eq.bound.cond.old}
\ee
where 
\be
C = \left( \frac{\tau_{0}'}{\tau_{0}} + A_{0} e^{\alpha^{0} \cdot \phi /2} \right).
\ee
Using this definition of $C$, the equations of motion imply, at the boundary $x=0$, 
\be
\ddot{\tau_{i}} - \frac{\dot{\tau_{i}}^{2}}{\tau_{i}} - \tau_{i}'' + 2 C \tau_{i}' - (C^2 - 1) \tau_{i} =0
\ee
which simplifies in the static case to
\be
\tau_{i}''-2C\tau_{i}' +(C^2-1) \tau_{i} |_{x=0}=0.
\ee
However this equation only contains the squares of the coefficients $A_{i}$ and so we no longer know which boundary conditions a given solution of this will obey. We will find that it is possible to recover this information later.

The parameter $C$ turns out to be proportional to the energy of the field by
\be
E = -2 \sum_{i=0}^{r} n_i C.
\ee

It is now desirable to try to find the lowest energy (i.e. highest $C$)
static background solutions to these equations. These will depend, in
general, on the boundary conditions imposed. It is known that there are
solutions to the equations which consist of solitonic solutions analytically
continued from the imaginary-coupling theory \cite{TH,B}. The tau functions for an $N$-soliton solution can be written as:
\be 
\tau_{j}(x,t) = \sum_{\mu_{1}=0}^{1} ... \sum_{\mu_{N}=0}^{1} \exp\left(\sum_{p=1}^{N} \mu_{p} (\Phi_{p}+\ln \omega^{a_{p} j}) + \sum_{1 \leq p \leq q \leq N} \mu_{p} \mu_{q} \ln A^{(a_{p} a_{q})}\right).
\ee
Here, $\Phi_{p} = \sigma_{p} (x - v_{p} t) + \xi_{p}$.
We have the mass-shell condition $\sigma^2 (1-v_{p}^2) = m_{a_{p}}^2$, which in the static case implies $\sigma=m_{a_{p}}$ since we need the asymptotic form $\tau_{i} \rightarrow 1$ as $x \rightarrow -\infty$. For $a_{r}^{(1)}$, the masses $m_{a_{p}}=2$sin$(\frac{\pi a_{p}}{r+1})$ and $\omega = $exp$(\frac{2 \pi i}{r+1})$. Later it is useful to take $\xi_{p}=\ln(d)+i \chi$ where $d$ is referred to as the position of the soliton.

The interactions between solitons are given by the interaction constants
\be
A^{(a_{p} a_{q})} = - \frac{(\sigma_{p} - \sigma_{q})^2 - (\sigma_{p} v_{p} -
\sigma_{q} v_{q})^2 - 4 \sin^2 \frac{\pi}{r+1} (a_{p} - a_{q})}
{(\sigma_{p} + \sigma_{q})^2 - (\sigma_{p} v_{p} + \sigma_{q} v_{q})^2
- 4 \sin^2 \frac{\pi}{r+1} (a_{p} + a_{q})}.
\ee
It is worth noting that when we interact one soliton with another of the same type and velocity, 
we find that the interaction constant $A^{(a_{p} a_{q})}$ vanishes. Hence in the static case it is only
possible to have at most one of each type of soliton in a background solution.

It is convenient to make the change of basis
\be 
\tau_{j}=\sum_{k} T_{k} \omega^{kj}
\ee
which splits the tau functions up into `charge sectors' $T_k$. This gives us
a similar equation to before but now for $T_{k}$:
\be 
T_{k}'' -2CT_{k}' + (C^{2} -1)T_{k} =0 
\label{chargesec}
\ee
at $x=0$. We now have a simpler set of equations: a soliton of type $a$ resides in $T_a$ whilst it's interaction with another soliton $b$ resides in $T_{a+b}$ and so on.

It is required that the function $\phi$ be real everywhere on the interval $(-\infty,0]$. Thus the tau functions must be real and non-negative on this interval. Reality imposes the condition that each soliton must appear with its conjugate, i.e. type $a$ with type $r+1-a$. The exception to this is where we have a middle soliton. The middle soliton has $a=\frac{r+1}{2}$ when $r$ is odd, and hence if we take $\chi=0,\pi$ we obtain a real tau function. However the restriction imposed by insisting that the tau function be non-negative is more subtle and will be considered case by case below. Let us look at possible background configurations in turn.

\subsection{$\phi=0$}

Looking at the boundary conditions (\ref{eq.bound.cond.old}) and remembering (\ref{sum.alphas}), it is clear that the solution $\phi=0$ is only valid when 
\be
0=\sum_{i=0}^{r} A_{i} \alpha^{i}.
\ee
This is clearly only true when all of the $A_{i}$ are the same sign. In
addition the value of $C$ is found to be $\pm 1$ for the $++..+$ and $--..-$
boundary conditions respectively. We use the notation for the boundary
conditions where the signs of the parameters $A_i$ are listed in turn.

\subsection{Single Middle Soliton}

In the case $r=\mbox{odd}$, there is a self-conjugate soliton associated with the central spot of the Dynkin diagram, i.e. of type $a=\frac{r+1}{2}$. If we consider a background configuration consisting only of such a static soliton, we find; 
\be
\tau_{j}=1+(-1)^{j} d e^{2 x}
\ee
where we have taken $\chi=0$ for simplicity ($\chi=\pi$ merely swaps the
r\^ole of the even and odd tau functions). It is apparent from this equation that we must take $d \leq 1$ in order that none of the tau functions are negative in the physical region. 

Since
\be 
e^{\alpha^{i} \cdot \phi} = \frac{\tau_{i-1} \tau_{i+1}}{\tau_{i}^2},
\ee
the boundary conditions imply that
\be
\frac{1-(-1)^j d e^{2x}}{1+(-1)^j d e^{2x}} = A_{j} \left| \frac{1-(-1)^j d e^{2x}}{1+(-1)^j d e^{2x}} \right|.
\ee
Note that here we have used the fact that the energy of such a solution is given by $C=m-1=1$. For $d < 1$, the modulus sign is
irrelevant and we obtain the result that all the $A_{i}$ must be positive. In
the special case $d=1$, all the odd tau functions vanish at the
boundary. Consider a vanishing tau function, $\tau_j$. Since we can write the $j-1$ and $j+1$ boundary conditions as
\be
C - \frac{\tau_{j-1}'}{\tau_{j-1}} = A_{j-1} \frac{1}{\tau_{j-1}} \sqrt{\tau_{j} \tau_{j-2}}
\ee
and
\be
C-\frac{\tau_{j+1}'}{\tau_{j+1}} = A_{j+1} \frac{1}{\tau_{j+1}} \sqrt{\tau_{j} \tau_{j+2}}
\ee
respectively, then it is clear that in this case we can take $A_{j-1}$ and $A_{j+1}$ to be of
arbitrary sign. This gives us a variety of consistent boundary conditions. It is
worth noting also that since $\tau_{j}' < 0$ and $\tau_{j} > 0$ for $x$ near
to but below $0$, then the boundary condition 
$A_{j} e^{\alpha_i \cdot \phi /2} = C - \frac{\tau_{j}'}{\tau_{j}} > 0$ 
for $x \rightarrow 0$ and so if $\tau_{j}$ vanishes at $x=0$ then $A_{j}=1$.
Hence when $d=1$, we can fit boundary conditions of the form $a+a+a+a+..$ where ``$a$'' denotes an arbitrary choice of sign.
Note that taking instead $\chi=\pi$ allows us to fit the boundary conditions of the form $+ \hspace{0.03in}a+a+a+..$.

\subsection{Two soliton solutions}

The single middle soliton was something of a special case: let us now consider the case where we have a soliton, type $a$, and its conjugate, type $\overline{a}=r+1-a$. Then the tau functions become;
\be
\tau_{i} = 1 + 2d \cos \left( \chi + \frac{2 \pi ia}{r+1} \right) e^{m_{a} x} + A^{(a \overline{a})} d^2 e^{2 m_{a} x}.
\ee
In this case it can readily be shown (using the equation for the highest
occupied charge sector) that there are two solutions $C_{\pm}=m_{a} \pm
1$. One can show that the lower energy solution $C_{+}$ has singularities in the physical region so we require the parity inverse solution $C_{-}$ \cite{B}. There is also a specific value of $d$ corresponding to this solution (determined by the zero charge sector) which is given by $d=\frac{2}{2+m_{a}}$.

Working with the boundary conditions (\ref{eq.bound.cond.old}) as before it can be shown that the coefficients $A_{i}$ are given by
\be
A_{i}=-\mbox{sign} \left( \cos \left( \frac{\chi}{2} + \frac{\pi (i+1)a}{r+1}
    \right) \cos \left( \frac{\chi}{2} + \frac{\pi (i-1) a}{r+1} \right) \right).
\ee
Note that a shift of the parameter $\chi$ by $2\pi / (r+1)$ merely cyclically 
permutes the boundary conditions obeyed. However, by choosing $\chi$ so that
$\cos (\frac{\chi}{2} + \frac{\pi ja}{r+1}) =0$ we can ensure that $\tau_{j}
\rightarrow 0$ as $x \rightarrow 0$. So as before this allows us to take the
signs of $A_{j-1}$ and $A_{j+1}$ to be arbitrary. Therefore there are a
number of boundary conditions consistent with a two-soliton solution which is
singular at the boundary.

\subsection {Multi-soliton solutions}

Let us first consider the case developed in \cite{B} where the highest charge
sector occupied $Q_{max} = \Sigma_p a_p \leq \frac{r+1}{2}$. We shall refer to
this case as having no {\em overlapping} of the charge-sectors, i.e. the
interaction term involving all the solitons $a_p$ always resides alone in
the $Q_{max}$ charge sector. This leads to the expressions for $C$ and $d_{p}$:
\be
C_{\pm}=\sum_{p=0}^{N} m_{a_{p}} \pm 1
\ee
\be 
d_{p}=\prod_{r \neq p} \frac{m_{a_{r}}+m_{a_{p}}}{|m_{a_{r}}-m_{a_{p}}|}
 \frac{2}{2 \mp m_{a_{p}}}
\ee
excepting that when $r$ is odd we are only allowed the $C_-$ solution.

Extensive numerical work with 3, 4, and 5 soliton solutions suggests that, under this regime, the only solutions which are non-zero in the region $(-\infty,0)$ occur when a maximal number of tau functions vanish at the boundary $x=0$. 
Consider a solution in which two or more consecutive tau functions vanish at
the origin. In the language of section three, this means that the $S$ contains at least
two neighbouring nodes. Let us rewrite the conditions (\ref{eq.sin.conditions1}) and (\ref{eq.sin.conditions}) for the
special case of $A_r^{(1)}$ theory. These give
\eqq
2y_i-y_{i+1}-y_{i-1} &=& 2 \label{cond1ar} \\
\frac{a_{i-1}a_{i+1}}{a_i^2} &=& y_i.
\label{cond2ar}
\enn
Now in the case where $S$ contains at least two consecutive nodes it follows
immediately from (\ref{cond1ar}) that $y_i$, the order of zero of
$\tau_i$ at $x=0$ must be greater than one.
However, using (\ref{cond1ar}) we see that for $i\in S$,
both sides of the boundary condition (\ref{eq.bound.cond.old}) 
\be 
C-\frac{\tau_i'}{\tau_i}=A_i \sqrt{\frac{\tau_{i-1} \tau_{i+1}}{\tau_i^2}}
\ee
go as $1/x$ as $x \rightarrow 0$. Hence, since $C$ (the energy) is finite,
the coefficients of these leading order terms must match. This requires,
using (\ref{cond2ar}),
\be
y_i = A_i \sqrt{y_i}
\label{singcond}
\ee
which cannot be satisfied for $y_i>1$. This tells us that only
non-consecutive tau functions tending to zero at the boundary are allowed. 
This result is important since numerical investigation suggests that, under
the non-overlapping charge sector regime, it is {\em not possible} to find
multi-soliton solutions which are both regular in the physical region {\em
  and} obey this restriction. However, a proof of this conjecture is lacking.

\subsubsection{Overlapping charge sectors, $Q_{max} > \frac{r+1}{2}$}

It is however sometimes possible to obtain acceptable multi-soliton solutions in cases where the charge sectors overlap. The first case where this is possible is $r=4$. Consider a solution containing all the possible solitons of the theory. Then look at the charge sectors:

\begin{center}
\begin{tabular}{cl}
Charge sector&Soliton combinations \\
-2&$\overline{2}$ $\overline{2}1\overline{1}$ 12 \\
-1&$\overline{1}$ $\overline{1}2\overline{2}$ $1 \overline{2}$ \\
0&0 $1 \overline{1}$ $2 \overline{2}$ $1 \overline{1} 2 \overline{2}$ \\
1&1 $12 \overline{2}$ $\overline{1} 2$\\
2&2 $21 \overline{1}$ $\overline{12}$ \\
\end{tabular}
\end{center}

Now we must apply the usual charge sector equations (\ref{chargesec}).
In the non-overlapping case, the highest occupied charge sector yields an
equation solely in $C$. Here, however, we obtain an equation in $d_{1}$ and $C$
which must be solved simultaneously with the other charge sector
equations. The three equations from (\ref{chargesec}) are:
\begin{eqnarray}
C^2-1 &+& d_{1}^2 A^{1 \overline{1}}[(C-2m_{1})^2 -1] + d_{2}^2 A^{2 \overline{2}} [(C-2m_{2})^2-1] \nonumber \\
 &+& d_{1}^2 d_{2}^2 (A^{12} A^{1 \overline{2}})^2 A^{1 \overline{1}} A^{2 \overline{2}} [(C-2(m_{1}+m_{2}))^2-1] =0
\end{eqnarray}
\begin{eqnarray}
(C-m_{1})^2-1 &+& d_{2}^2 A^{12} A^{1 \overline{2}} A^{2 \overline{2}} [(C-(m_{1}+2m_{2}))^2-1] \nonumber \\
&+& d_{2} e^{i(\chi_{2}-2\chi_{1})} A^{1 \overline{2}} [(C-(m_{1}+m_{2}))^2-1] =0 
\end{eqnarray}
and
\begin{eqnarray}
(C-m_{2})^2-1 &+& d_{1}^2 A^{12} A^{1 \overline{2}} A^{1 \overline{1}} [(C-(m_{2}+2m_{1}))^2-1] \nonumber \\
&+& d_{1} e^{-i(\chi_{1}+2\chi_{2})} A^{12} [(C-(m_{1}+m_{2}))^2-1]=0.
\end{eqnarray}

If we take $e^{i(\chi_{2}-2\chi_{1})}$ and $e^{-i(\chi_{1}+2\chi_{2})}$ to be real we can find another real solution to these equations. We shall call this the ``exceptional'' solution. This has energy $C=m_{1}+m_{2} - \sqrt{5}$ and we can once again determine the boundary conditions which it obeys. What is remarkable is that in this case, this exceptional solution is regular in the physical region, has non-consecutive zeroes of the tau functions at $x=0$, and fits exactly those boundary conditions not covered by the non-exceptional solutions.

We can also use this technique for the case $r=5$ where, despite greater difficulties with finding solutions to the four simultaneous equations, we obtain similar results.

\subsection{Vacuum Solution Results}

Before presenting our results for the static background solutions for the
$A_{r}^{(1)}$ series of affine Toda field theories up to $r=5$ below, let us
quickly recap the allowed background solutions. Of the non-exceptional
solutions, there are few allowed background configurations. In the case
$r=\mbox{even}$, the only allowed background solutions contain either no
solitons or a soliton/conjugate soliton pair. When $r=\mbox{odd}$ we are also
allowed the case of a single middle soliton. In any of these cases, the
singularity may reside at the boundary $x=0$, allowing additional choices in
the boundary conditions obeyed. However, when $r \ge 4$, these configurations do
not span all the possible boundary conditions. To find the vacuum solutions
for these remaining boundary conditions, it is necessary to
consider the ``exceptional'' solutions. In the cases considered these
exceptional solutions completely cover the boundary conditions not covered by the non-exceptional
configurations. It is expected that similar results will extend beyond $r=5$
although a proof is lacking. 

It should be noted that, in many cases, more than one possible background
configuration fits with a particular boundary condition. When this occurs,
the vacuum solution is of course that background configuration which has lowest energy.

Our results up to $r=5$ are tabulated in table \ref{tabback}. These
results are ordered within each theory with decreasing $C$, or equivalently,
increasing energy.

\begin{table}

\begin{center}

\begin{tabular}{|c|c|c|c|} \hline
Boundary Condition & Solitons & $C$
& Singular at $x=0$ \\ \hline
$+++$ & none & 1 & no \\ \hline
\( \begin{array}{c} -++ \\ --+ \end{array} \) & $1 \overline{1}$ & $\sqrt{3} -1$ & \( \begin{array}{c} \mbox{no} \\ \mbox{yes} \end{array} \) \\ \hline
$---$ & none & -1 & no \\ \hline
\end{tabular}
\vspace{0.05in}

$r=2$ 

\vspace{0.1in}

\begin{tabular}{|c|c|c|c|} \hline
Boundary Condition & Solitons & $C$
& Singular at $x=0$ \\ \hline
$++++$ & none & 1 & no \\ \hline
\( \begin{array}{c} -+++ \\ -+-+ \end{array} \) & $2$ & $1$ & \( \begin{array}{c} \mbox{yes} \\ \mbox{yes} \end{array} \) \\ \hline
\( \begin{array}{c} --++ \\ ---+ \end{array} \) & $1 \overline{1}$ & $\sqrt{2} -1$ & \( \begin{array}{c} \mbox{no} \\ \mbox{yes} \end{array} \) \\ \hline
$----$ & none & -1 & no \\ \hline
\end{tabular} 
\vspace{0.05in}

$r=3$ 

\vspace{0.1in}

\begin{tabular}{|c|c|c|c|} \hline
Boundary Condition & Solitons & $C$
& Singular at $x=0$ \\ \hline
$+++++$ & none & 1 & no \\ \hline
\( \begin{array}{c} -++++ \\ -+-++ \end{array} \) & $2 \overline{2}$ & $\frac{1}{\sqrt{2}} \sqrt{5+\sqrt{5}} -1$ & \( \begin{array}{c} \mbox{no} \\ \mbox{yes} \end{array} \) \\ \hline
\( \begin{array}{c} --+++ \\ --+-+ \end{array} \) & $1 \overline{1} 2 \overline{2}$ & $\sqrt{5+2\sqrt{5}} - \sqrt{5}$ & \( \begin{array}{c} \mbox{yes} \\ \mbox{yes} \end{array} \) \\ \hline
\( \begin{array}{c} ---++ \\ ----+ \end{array} \) & $1 \overline{1}$ & $\frac{1}{\sqrt{2}} \sqrt{5-\sqrt{5}} -1$ & \( \begin{array}{c} \mbox{no} \\ \mbox{yes} \end{array} \) \\ \hline
$-----$ & none & -1 & no \\ \hline
\end{tabular} 
\vspace{0.05in}

$r=4$

\vspace{0.1in}

\begin{tabular}{|c|c|c|c|} \hline
Boundary Condition & Solitons & $C$
& Singular at $x=0$ \\ \hline
$++++++$ & none & 1 & no \\ \hline
$-+++++$ & & & yes \\  
$-+-+++$ & 3 & 1 & yes \\ 
$-+-+-+$ & & & yes \\ \hline
\( \begin{array}{c} -++-++ \\ --++++ \\ --++-+ \\ --+-++ \\ --+--+
\end{array} \) & $2 \overline{2}$ & $\sqrt{3} -1$ & \( \begin{array}{c}
  \mbox{no} \\ \mbox{yes} \\ \mbox{yes} \\ \mbox{yes} \\ \mbox{yes} \end{array} \) \\ \hline
\( \begin{array}{c} ---+++ \\ ---+-+ \end{array} \) & $1 \overline{1} 2 \overline{2} 3$ & $\sqrt{3} + 2 - \sqrt{2(2^{1/3}+2^{2/3}+2)}$ & \( \begin{array}{c} \mbox{yes} \\ \mbox{yes} \end{array} \) \\ \hline
\( \begin{array}{c} ----++ \\ -----+ \end{array} \) & $1 \overline{1}$ & $0$ & \( \begin{array}{c} \mbox{no} \\ \mbox{yes} \end{array} \) \\ \hline
$------$ & none & -1 & no \\ \hline
\end{tabular} 
\vspace{0.05in}

$r=5$ 
\vspace{-0.05in}
\caption{Constituent solitons and energy parameters $C$ for the vacuum solutions of $A_r^{(1)}$ affine Toda field theory on a half-line
with integrable boundary conditions, given up to $r=5$. \label{tabback}}

\end{center}

\end{table}

\section{Classical Scattering Solutions}

Besides classical backgrounds, more information can be obtained from the classical equations of motion. The classical limit of the reflection matrix 
$K_a$ can be found by considering solutions which correspond
to infinitesimal classical fluctuations around the background configuration 
\cite{CDR}.

Suppose the field is perturbed about the vacuum
\be
\phi (x,t) = \phi_{vac}(x) + \epsilon(x,t).
\ee
Then by considering the equations of motion we find that, in the limit $ x
\rightarrow -\infty$, $\epsilon (x,t)$ may have the asymptotic form
\be
\epsilon(x,t) \rightarrow \rho_{a} e^{-iEt} (e^{ipx} + K_{a} e^{-ipx}),
\ee 
consisting of two superposed incoming and outgoing states. Here $\rho_a$ is a mass-eigenstate; an eigenvector of the mass matrix $M^2=\Sigma_i \alpha^i \otimes \alpha^i$. Solutions of this form which obey the boundary conditions can be generated by considering non-static two soliton solutions, where now 
\be
\Phi(x,t) = \sigma (x-vt)
\ee
is defined by 
\be
\sigma = \pm ip
\ee
\be
\sigma v = iE.
\ee
Hence we must add these two solitons both of type $a$ to the background configuration and determine $K_{a}$ by insisting that the resulting tau functions satisfy
\be
\ddot \tau_{i} - \tau_{i}'' + 2C \tau_{i}' - (C^2-1) \tau_{i} |_{x=0} =0
\ee
or of course equivalently from the charge sector point of view
\be
\ddot T_{i} - T_{i}'' + 2C T_{i}' - (C^2-1) T_{i} |_{x=0} =0.
\ee
It is then very simple to determine the values of $K$ corresponding to different boundary conditions.

Although they have already been set out in \cite{B}, let us briefly restate
the results for the classical reflection factors in the cases which have
non-exceptional vacuum solutions. In fact, we shall see that this is not necessarily
relevant to cases which are singular at the boundary $x=0$. Solving the equation for the highest occupied charge sector reveals that
\be
K_a=\frac{2ip+m_a^2}{2ip-m_a^2} \prod_{j=1}^{N} \frac{A^{ab_j}(p)}{A^{ab_j}(-p)}
\ee
where $a$ is the particle being scattered and $b_j$ are the solitons present in the background. $N$ refers to the number of soliton species (not including conjugates) present in the background.
The only exception to this is the flat background case $++...+$, where
\be
K_a=\frac{2ip-m_a^2}{2ip+m_a^2}.
\ee
It has been checked in \cite{B} that these reflection factors obey the classical reflection bootstrap equation.

There is however a little more subtlety involved when we consider the cases
where one or more of the background tau functions vanish at $x=0$. In this
case it is not so clear that a linear perturbation in $\phi \rightarrow \phi
+ \epsilon$ and in the tau functions $\tau_i \rightarrow \tau_i + \epsilon_i$
are equivalent. However, with certain restrictions on the $\epsilon_i$
(namely that if $\tau_i \rightarrow 0$ as $x \rightarrow 0$ then $\epsilon_i
\rightarrow 0$ as quickly or faster in order to result in a finite
perturbation $\epsilon$) we can achieve the same results. Since the boundary
conditions tell us that from the tau function point of view
\be
C - \frac{\tau_{i}'}{\tau_i} = A_{i} \sqrt{\frac{\tau_{i-1} \tau_{i+1}}{\tau_i^2}}
\ee
then if we take $i$ so that $\tau_{i+1}$ goes to zero whilst $\tau_{i-1}$ and
$\tau_{i}$ do not, then clearly we require the term linear in $\epsilon$ in
$\tau_{i+1}$ to go to zero at $x=0$. In fact, this is almost the same
restriction as the requirement that the perturbation $\epsilon$ is
finite. This condition gives us another equation which must be satisfied by
the reflection factor $K$. It is simplest to consider the case where we take
the parameters $e^{i \chi}$ to be real, so that the vacuum solution tau
functions are symmetric about $\tau_0$ (i.e. $\tau_{r-i+1}=\tau_i$) --- the
other boundary conditions are of course related by the cyclical symmetry. Let
us illustrate our point by considering the case of $A_2^{(1)}$ for the
boundary conditions $+--$ or $++-$ which contains two static solitons in
the background solution. Here, let us take $\chi=\pi$ so that, as far as the
background solution is concerned, $\tau_0 \rightarrow 0$ as $x \rightarrow 0$
and $\tau_1=\tau_2$. Now adding in the perturbation, we require the right-hand sides of the two boundary conditions
\be
C - \frac{\tau_{1}'}{\tau_1} = A_{1} \sqrt{\frac{\tau_{0} \tau_{2}}{\tau_1^2}}
\ee
and
\be
C - \frac{\tau_{2}'}{\tau_2} = A_{2} \sqrt{\frac{\tau_{0} \tau_{1}}{\tau_2^2}}
\ee
(evaluated at $x=0$) to be equal up to $O(\epsilon)$, with a possible sign difference between $A_1$ and $A_2$. Hence we expect 
\be
A_1 \frac{\tau_{1}'}{\tau_1} = A_2 \frac{\tau_{2}'}{\tau_2}
\label{nondiagcond}
\ee
up to $O(\epsilon)$.
In actual fact, for $A_1=-A_2$, this equation is satisfied identically for
all $K$ and hence the same scattering data as we found before for the
non-singular background also obeys this boundary condition. However, for
$A_1=A_2$ we require both sides of the equation to vanish. Imposing this
condition leads to a somewhat unaesthetic non-diagonal scattering solution
also found by Corrigan \cite{CP}. Diagonal scattering is expected in
$A_r^{(1)}$ as the conservation of spin even charges (which distinguish
between particles and their conjugates) implies that a particle should
be reflected back into itself \cite{CDR}, and is moreover implied by the 
bootstrap equations. To see this, consider the bootstrap equation depicted
in Figure 1 for the
non-diagonal case. Consider the case where $c=1$. Then non-diagonal scattering
implies that $c$ can be reflected as a particle of type 1 or 2. This particle will then
decompose into two particles $a$ and $b$, either of type 1 or 2, but in any
case $a=b$. However, the RHS implies that $c$ will split first into two
particles of type 2, which are then each reflected as types 1 or 2. In this
case there is no requirement that $a=b$. Hence the scattering must be either diagonal (as
generally believed for real coupling Toda theory) or completely off-diagonal
(a possibility suggested by the recent paper by Delius \cite{D2}). This is
rules out the unaesthetic non-diagonal scattering solution for the case
$A_1=A_2$.

\begin{figure}
\begin{center}
\epsfig{figure=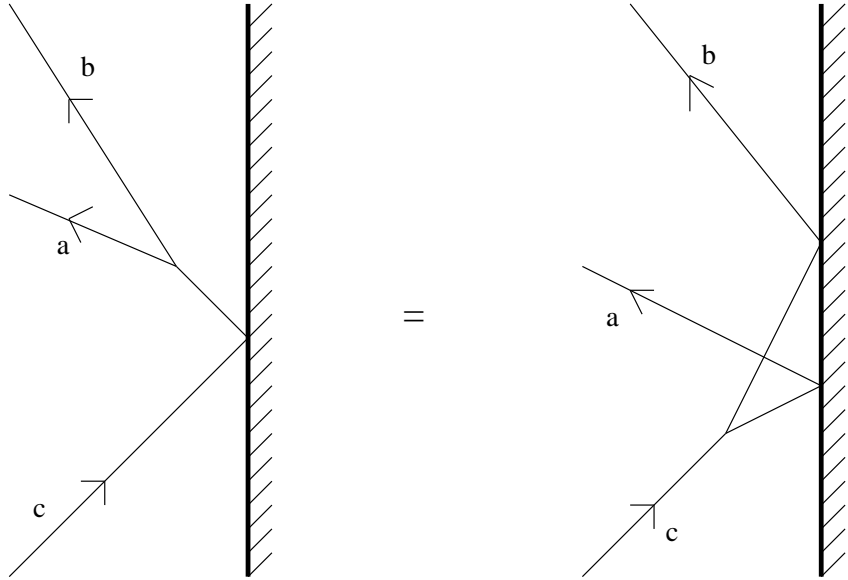}
\caption{The boundary bootstrap equation.
\label{fig:refboot}}
\end{center}
\end{figure}

Hence we find that there is no consistent classical scattering data for the boundary condition $+--$. 

It is not difficult to extend the above argument to include {\em all}
vacuum solutions in $A_r^{(1)}$ affine Toda theory consisting of two static solitons. When
the singularity is placed at the boundary it can be shown that, for any $i
\in S$, we obtain an analogous equation to (\ref{nondiagcond}), involving the
boundary parameters $A_{i-1}$ and $A_{i+1}$. This relation is only solved for
diagonal scattering if we takes these two parameters to be opposite. This rules out a large subset of
the integrable boundary conditions. Whilst it would appear that there is no
problem with singular background solutions containing only one static middle
soliton (here, the signs of the (say) odd $i$ boundary parameters remain
arbitrary since the $O(\epsilon)$ term of $\frac{\tau'_i}{\tau_i}$ vanishes),
we also find that the `exceptional' solutions of $A_4^{(1)}$ and $A_5^{(1)}$
have problems. For example, using the exceptional solution of $A_4^{(1)}$, we
find that there are two vanishing tau functions, $\tau_1$ and $\tau_4$
(preserving the symmetry around $\tau_0$). Then, as noted earlier, we require the linear
perturbations of both of these vanishing tau functions to vanish at
$x=0$. But each of these requires a different $K$. A similar conclusion is
reached when considering the $A_5^{(1)}$ exceptional solution. So in these cases, it seems that no diagonal scattering solutions
are possible which go as $\rho_a$ as $x \rightarrow - \infty$. It would
therefore appear that a large subset of the integrable boundary conditions do
not admit suitable classical scattering solutions. The ones that do fall into
one of three categories: they are consistent with a background configuration which either contains no solitons, a single middle
soliton (in the case $r=\mbox{odd}$) or a soliton/conjugate soliton pair
where the signs of the boundary parameters on each side of a vanishing tau
function are opposite.

\subsection{Classical Scattering Results}

We use the usual notation for expression of the reflection factors. We define
\be
(x)=\frac{\sinh(\frac{\theta}{2} + \frac{i \pi x}{2(r+1)})}{\sinh(\frac{\theta}{2} - \frac{i \pi x}{2(r+1)})}.
\ee

The results for the reflection factors for $r$ up to 5 are given in table
\ref{tabscat}.

\begin{table}

\begin{center}

\begin{tabular}{|c|c|} \hline
Boundary Condition & $K_{1}$, $K_{\overline{1}}$ \\ \hline
$+++$ & $-(1)(2)$ \\ \hline
$-++$ & $-\frac{(\frac{1}{2})(\frac{3}{2})^2 (\frac{5}{2})}{(1)(2)}$ \\ \hline 
$--+$ & not diagonal\\ \hline
$---$ & $-\frac{1}{(1)(2)}$ \\ \hline
\end{tabular}
\vspace{0.03in}    

$r=2$ 

\vspace{0.05in}

\begin{tabular}{|c|c|c|} \hline
Boundary Condition & $K_{1}$, $K_{\overline{1}}$ & $K_{2}$ \\ \hline
$++++$ & & \\ 
$-+++$ & $-(1)(3)$ & $-(2)(2)$ \\  
$-+-+$ & & \\ \hline
$--++$ & $\frac{(2)^2}{(1)(3)}$ & $-\frac{(1)^2 (3)^2}{(2)^2}$ \\ \hline 
$---+$ & not diagonal & not diagonal \\ \hline
$----$ & $-\frac{1}{(1)(3)}$ & $-\frac{1}{(2)^2}$ \\ \hline
\end{tabular} 
\vspace{0.03in}

$r=3$ 

\vspace{0.05in}

\begin{tabular}{|c|c|c|} \hline
Boundary Condition &  $K_{1}$, $K_{\overline{1}}$ &  $K_{2}$, $K_{\overline{2}}$  \\ \hline
$+++++$ & $-(1)(4)$ & $-(2)(3)$ \\ \hline
$-++++$ & $-\frac{(\frac{1}{2})(\frac{3}{2})(\frac{7}{2})(\frac{9}{2})}{(1)(4)}$ & $-\frac{(\frac{3}{2})(\frac{5}{2})^2 (\frac{7}{2})}{(2)(3)}$ \\ \hline 
$-+-++$ & not diagonal & not diagonal \\ \hline
\( \begin{array}{c} --+++ \\ --+-+ \end{array} \) & not diagonal & not diagonal \\ \hline
$---++$ & $-\frac{(\frac{5}{2})^2}{(\frac{1}{2})(1)(4)(\frac{9}{2})}$ & $-\frac{(\frac{1}{2})(\frac{3}{2})(\frac{7}{2})(\frac{9}{2})}{(2)(3)}$ \\ \hline
$----+$ & not diagonal & not diagonal \\ \hline
$-----$ & $-\frac{1}{(1)(4)}$ & $-\frac{1}{(2)(3)}$ \\ \hline
\end{tabular} 
\vspace{0.03in}

$r=4$ 

\vspace{0.05in}

\begin{tabular}{|c|c|c|c|} \hline
Boundary Condition & $K_1$, $K_{\overline{1}}$  & $K_2$, $K_{\overline{2}}$ & $K_3$ \\ \hline
\( \begin{array}{c} ++++++ \\ -+++++ \\ -+-+++ \\ -+-+-+ \end{array} \) & $-(1)(5)$ & $-(2)(4)$ & $-(3)^2$ \\ \hline
\( \begin{array}{c} -++-++ \\ --++++ \end{array} \) & $\frac{(2)(4)}{(1)(5)}$ & $-\frac{(1)(3)^2(5)}{(2)(4)}$ & $-\frac{(2)^2(4)^2}{(3)^2}$ \\ \hline
\( \begin{array}{c} --++-+ \\ --+-++ \\ --+--+ \end{array} \) & not diagonal & not diagonal & not diagonal \\ \hline
\( \begin{array}{c} ---+++ \\ ---+-+ \end{array} \) & not diagonal & not diagonal & not diagonal \\ \hline
$----++$ & $-\frac{(3)^2}{(1)^2(5)^2}$ & $1$ & $-\frac{(1)^2(5)^2}{(3)^2}$ \\ \hline
$-----+$ & not diagonal & not diagonal & not diagonal \\ \hline 
$------$ & $-\frac{1}{(1)(5)}$ & $-\frac{1}{(2)(4)}$ & $-\frac{1}{(3)^2}$ \\ \hline
\end{tabular} 

\vspace{0.03in}

$r=5$

\vspace{-0.05in}

\caption{Classical reflection factors for $A_r^{(1)}$ affine Toda field
theory with integrable boundary conditions, given up to $r=5$. \label{tabscat}}

\end{center}

\end{table}

\section{Conclusions}

In this paper we have presented a number of results about classical solutions to Affine Toda field theory on a half-line. Extending what had previously been achieved for the special case of sinh-Gordon \cite{CDR},
we used a Bogomolny argument to ascertain which boundary conditions lead to stable theories with bounded
energy on the half-line. This class includes all integrable boundary conditions for simply-laced affine Toda theories, although many lie on the boundary of
stability. 

However, it turns out that there are still difficulties with providing
a classical picture of scattering in some cases. This was found by considering the vacuum and scattering solutions for low rank $A_r^{(1)}$ affine Toda field theories in great detail
in order to attempt to shed some light on what happens in general. We have
found many unusual characteristics of these theories. It was found that
although acceptable vacuum solutions can be found for all the ``integrable''
boundary conditions, not all of these appear to admit classical scattering
data consistent with the reflection bootstrap equation. Another problem
is that many of the classical background solutions contain parameters or 
moduli describing equally valid degenerate ground states. Exact solutions which
correspond to excitations of these `zero-modes' can be found, but they 
are problematic in that they become singular in some finite time. If 
analytically continued beyond this time the solutions then satisfy changed boundary conditions. 

Finally, it would be
interesting to compare these results with those of \cite{GAND,DG} where
quantum reflection factors for $A_r^{(1)}$ ATFT were considered.

\section{Acknowledgements}

The authors would like to thank E. Corrigan, P. Dorey and G. Delius for enlightening discussions on this topic. MGP also acknowledges the support of the UK Engineering and Physical Sciences Research Council.

\end{document}